\documentclass[fleqn,11pt,twoside]{article}
\usepackage{amsthm,amsthm,amssymb, color, epsfig, graphics, subfigure,}
\usepackage{tikz}
\usepackage{amsmath}%[fleqn,intlimits]
\usepackage{graphicx}
\usepackage{amsmath}%[fleqn,intlimits]
\usepackage{graphicx}
\usepackage{latexsym}
\usepackage[all]{xy}
\usepackage{amsthm}
\usepackage{amsmath, amscd}%[fleqn,intlimits]
\usepackage{graphicx}
\usepackage{lscape}
\usepackage{amsmath}%[fleqn,intlimits]
\usepackage{graphicx}
\usepackage{latexsym}
\newtheorem{proposition}{Proposition}
\newtheorem{remark}{Remark}

\makeatletter
\newcommand{\copyrightnote}[2]{{\renewcommand{\thefootnote}{}
 \footnotetext{\small\it
\begin{flushleft}
\copyright \ #1   #2
\end{flushleft}}}}

\newcommand{\Name}[1]{\begin{flushleft}
                       \LARGE \bf #1
                       \end{flushleft}\vspace{-3mm}}

\newcommand{\Author}[1]{\begin{flushleft}
                       \it #1 \end{flushleft}}

\newcommand{\Address}[1]{\begin{flushleft}
                       \it #1 \end{flushleft}}

\newcommand{\evenhead}{Author \ name}
\newcommand{\oddhead}{Article \ name}

\renewcommand{\@evenhead}{
\hspace*{-3pt}\raisebox{-15pt}[\headheight][0pt]{\vbox{\hbox to \textwidth
{\thepage \hfil \evenhead}\vskip4pt \hrule}}}
\renewcommand{\@oddhead}{
\hspace*{-3pt}\raisebox{-15pt}[\headheight][0pt]{\vbox{\hbox to \textwidth
{\oddhead \hfil \thepage}\vskip4pt\hrule}}}
\renewcommand{\@evenfoot}{}
\renewcommand{\@oddfoot}{}

\long\def\@makecaption#1#2{%
  \vskip\abovecaptionskip
  \sbox\@tempboxa{\small \textbf{#1.}\ \ #2}%
  \ifdim \wd\@tempboxa >\hsize
    {\small \textbf{#1.}\ \ #2}\par
  \else
    \global \@minipagefalse
    \hb@xt@\hsize{\hfil\box\@tempboxa\hfil}%
  \fi
  \vskip\belowcaptionskip}

\newcommand{\JNMPnumberwithin}[3][\arabic]{%
  \@ifundefined{c@#2}{\@nocounterr{#2}}{%
    \@ifundefined{c@#3}{\@nocnterr{#3}}{%
      \@addtoreset{#2}{#3}%
      \@xp\xdef\csname the#2\endcsname{%
        \@xp\@nx\csname the#3\endcsname .\@nx#1{#2}}}}%
}

\newcommand{\resetfootnoterule} {
  \renewcommand\footnoterule{%
  \kern-3\p@
  \hrule\@width.4\columnwidth
  \kern2.6\p@}
}

\renewcommand{\footnoterule}{}

\newcommand{\be}{\begin{equation}}
\newcommand{\ee}{\end{equation}}
\newcommand{\ba}{\hspace*{-5pt}\begin{array}}
\newcommand{\ea}{\end{array}}
\newcommand{\p}{\partial}

\makeatother
\numberwithin{equation}{section}
\theoremstyle{definition}

\renewcommand{\ba}{\begin{array}}
\renewcommand{\ea}{\end{array}}
\newcommand{\beg}{\begin{eqnarray}}
\newcommand{\eeq}{\end{eqnarray}}
\newcommand{\bg}{\begin{eqnarray*}}

\newcommand{\ed}{\end{eqnarray*}}

\newcommand{\nn}{\nonumber}

\renewcommand{\p}{\partial} 
 
\newcommand{\notlhd}{\lhd\kern-.8em{/}\ } 
\newcommand{\notexist}{\ \exists\kern-.5em{\raise.1em\hbox{/}}\ }

\newcommand{\pde}[2]{\frac{\p #1}{\p #2}}

\newcommand{\inp}{{\mbox{\vbox{\hrule width0ex\hbox{\vrule
 height0ex\kern3.8pt
\vbox{\kern2.5pt}\kern3.8pt \vrule height1.6ex}
\hrule width1.6ex}}}}

\begin{document}

%\Large

\renewcommand{\evenhead}{ {\LARGE\textcolor{blue!10!black!40!green}{{\sf \ \ \ ]ocnmp[}}}\strut\hfill M. Euler and N. Euler}
\renewcommand{\oddhead}{ {\LARGE\textcolor{blue!10!black!40!green}{{\sf ]ocnmp[}}}\ \ \ \ \ On fully-nonlinear 3rd-order symmetry-integrable equations}

% Title

\thispagestyle{empty}

\newcommand{\FistPageHead}[3]{
\begin{flushleft}
\raisebox{8mm}[0pt][0pt]
{\footnotesize \sf
\parbox{150mm}{{Open Communications in Nonlinear Mathematical Physics}\ \ \ {\LARGE\textcolor{blue!10!black!40!green}{]ocnmp[}}
\ \ Vol.2 (2022) pp
#2\hfill {\sc #3}}}\vspace{-13mm}
\end{flushleft}}

\FistPageHead{1}{\pageref{firstpage}--\pageref{lastpage}}{ \ \ Letter}

\strut\hfill

\strut\hfill

\copyrightnote{The author(s). Distributed under a Creative Commons Attribution 4.0 International License}

\qquad\qquad\qquad\qquad\qquad\qquad {\LARGE  {\sf Letter to the Editors}}

\strut\hfill

\Name{On fully-nonlinear symmetry-integrable equations with rational functions in their highest derivative: Recursion operators}

\label{firstpage}

%\strut\hfill

%\strut\hfill

%\strut\hfill

\Author{Marianna Euler and Norbert Euler$^{\, *}$ }

%\strut\hfill

%\strut\hfill

\Address{
Centro Internacional de Ciencias, Av. Universidad s/n, Colonia Chamilpa,\\
 62210 Cuernavaca, Morelos, Mexico\\
 $^*$ Corresponding author: Dr.Norbert.Euler@gmail.com}

%\vspace{1cm}

%\vspace{1cm}

%\strut\vfill

%\pagebreak

\noindent
{\bf Abstract}: 
We report a class of symmetry-intergable third-order evolution equations in 1+1 dimensions under the condition that the equations admit a second-order recursion operator that contains an adjoint symmetry (integrating factor) of order six. The recursion operators are given explicitly. 

\strut\hfill

%MSC-class: 37K35, 35C05

%%%%%%%%%%%%%%%%%%%%%%%%

\renewcommand{\theequation}{\arabic{section}.\arabic{equation}}

\allowdisplaybreaks

\section{Introduction}

We recently reported four fully-nonlinear Möbius-invariant and symmetry-integrable third-order evolution equations, namely \cite{E-E-April2019}
\begin{subequations}
\begin{gather}
\label{S-Eq-1}
u_t=\frac{u_x}{(b-S)^2},\quad b\neq 0\\[0.3cm]
\label{S-Eq-2}
u_t=\frac{u_x}{S^2}\\[0.3cm]
\label{S-Eq-3}
u_t=-2\frac{u_x}{\sqrt{S}}\\[0.3cm]
\label{S-Eq-4}
u_t=\frac{u_x(a_1-S)}{(a_1^2+3a_2)(S^2-2a_1S-3a_2)^{1/2}},\quad a_1^2+3a_2\neq 0,
\end{gather}
\end{subequations}
where $S$ denotes the Schwarzian Derivative
\begin{gather}
\label{Schwarzian}
S:=\frac{u_{xxx}}{u_x}-\frac{3}{2} \left(\frac{u_{xx}}{u_x}\right)^2.
\end{gather}
This classification was achieved by matching quasi-linear auxiliary symmetry-integrable evolution equations in $S$ for each equation 
(\ref{S-Eq-1}) -- (\ref{S-Eq-4}).
%, which therefore did not require us to compute recursion operators for (\ref{S-Eq-1}) -- (\ref{S-Eq-4}). 
%See \cite{E-E-April2019} for details. 
In \cite{E-E-JNMP2021} we propose a method to compute the higher members of the hierarchies of (\ref{S-Eq-1}) -- (\ref{S-Eq-4}) without the knowledge of the equations' recursion operators. 
%It is importasnt to note that this proposed method does not require one to compute a recursion operator for the fully nonlinear equations (\ref{S-Eq-1}) -- (\ref{S-Eq-1}). 
In particular, the proposed method makes use of the recursion operators of the 
auxiliary quasi-linear evolution equations in the variable $S$. This is an essential point since it is in general rather complicated and tedious to compute recursion operators, especially for fully-nonlinear equations. It is important to point out that the method to compute the higher-order members of the hierarchies as proposed in \cite{E-E-JNMP2021}, only applies to evolution equations that are Möbius-invariant and symmetry-integrable. Furthermore we point out that it is not possible to extend the idea of Möbius-invariant evolution equation to systems of evolution equations in a direct sense. This has been investigated in \cite{EulerEulerNucci2022}.

%\strut\hfill
%
%
%{\remark 
%Equations (\ref{S-Eq-1}) and (\ref{S-Eq-2}) have an essentially different higher-order symmetry structure and it is not possible to set $b=0$ 
%in the hierarchy for (\ref{S-Eq-1}) to obtain the hierarchy for (\ref{S-Eq-2}). Therefore we list these two equations as separate cases.
%}

Inspired by the above mentioned results, we address here the problem of identifying fully-nonlinear symmetry-integrable evolution equations 
beyond the Möbius-invariant class and we do so by requiring the equations to admit a recursion operator of a certain form. In particular, we restrict ourselves to evolution equations that contain rational functions in $u_{xxx}$. Moreover, we assume a recursion operator of order two with an integrating factor of maximum order six. 
This of course restricts us to a special class of equations, namely equations that admit those type of recursion operators. Nevertheless, we believe that our findings are of interest and that the results reported here are new. 
%We aim to extend the study further to include algebraic functions in $u_{xxx}$ in a future project. 
%The Schwarzian equations (\ref{S-Eq-1}) and 
%(\ref{S-Eq-2}) are included in our findings.

We would like to point out that Hern\'andez Heredero \cite{H-H-2005} classified a type of
third-order integrable fully-nonlinear evolution equations that does not include equations with rational functions in $u_{xxx}$.
% and which is therefore outside the scope of our paper. 

%In addition, we also make use of the integrating factors of order four (zero-order and second-order do not exist for the obtained equations)
%and compute all the potentialisations of our fully nonlinear equations.

\section{Notations and conditions}

To fix the notation and to recall the conditions that are needed in this paper, we consider the general $n$th-order autonomous evolution equation in 1+1 dimensions
\begin{gather}
\label{G-Eeq-1}
E:=u_t-F(u,u_x,u_{xx},u_{xxx},\ldots,u_{nx})=0.
\end{gather}
The subscripts of $u$ denote partial derivatives, where partial derivatives of order 4 and higher are indicated by $u_{nx}$, $n\geq 4$.

Equation  (\ref{G-Eeq-1}) is said to be {\it symmetry-integrable} if it admits a recursion operator $R[u]$ that generates an infinite number of local Lie-Bäcklund (or generalized) symmetries for the equation. In this paper we consider recursion operators of the following form
\begin{gather}
\label{R-nth-order}
R[u]:= 
\sum_{k=1}^mG_k[u]D_x^k +G_0[u]+\sum_{j=1}^p I_j[u]D_x^{-1}\circ \Lambda_j[u].
\end{gather}
The notation $R[u]$ and $G_j[u]$ indicates that the operator $R$ and functions $G_j$
depend on $u,\ u_x,\ u_{xx},\ldots$ up to an order that is {\it ab initio} not fixed.
Here $I_j$ are Lie-B\"acklund symmetry coefficients for (\ref{G-Eeq-1}), i.e.
the coefficients of a symmetry generator
\begin{gather}
Z=I_j[u]\pde{\ }{u}
\end{gather}
which satisfies the condition
\begin{gather}
\left.\vphantom{\frac{DA}{DB}}
L_E[u]I_j[u]\right|_{E=0}=0, 
\end{gather}
where $L_E[u]$ denoted the linear operator
\begin{gather}
L_E[u]:=\pde{E}{u}+\pde{E}{u_t}D_t+\pde{E}{u_x}D_x+\pde{E}{u_{xx}}D_x^2+\cdots
+\pde{E}{u_{nx}}D_x^n.
\end{gather}
$\Lambda_j[u]$ are integrating factors for conservation laws 
\begin{gather}
\label{conservation}
\left.\vphantom{\frac{DA}{DB}}
D_t\Phi^t[u]+D_x\Phi^x[u]\right|_{E=0}=0,
\end{gather}
of (\ref{G-Eeq-1}), where
\begin{gather}
\Lambda[u]=\hat E[u] \Phi^t[u]
\end{gather}
and $\Lambda$ must satisfy the condition
\begin{gather}
\label{E-hat-cond-1}
\left.
\vphantom{\frac{DA}{DB}}
\hat E[u]\left( \Lambda[u]E\right)\right|_{E=0}=0.
\end{gather}
Here $\hat E[u]$ is the Euler operator
\begin{gather}
\hat E[u]:=\pde{\ }{u}-D_t\circ \pde{\ }{u_t}-D_x\circ \pde{\ }{u_x}
+D_x^2\circ \pde{\ }{u_{xx}}-D_x^3\circ \pde{\ }{u_{3x}}+\cdots.
\end{gather}
Note that condition (\ref{E-hat-cond-1}) is equivalent to
\begin{subequations}
\begin{gather}
\label{ajoint-symmetry}
\left.\vphantom{\frac{DA}{DB}}
L^*_E[u] \Lambda[u]\right|_{E=0}=0\\[0.3cm]
\label{self-adjoint}
\mbox{and}\quad 
L_\Lambda[u]E=L_\Lambda^*[u]E.
\end{gather}
\end{subequations}
The first condition (\ref{ajoint-symmetry}) requires $\Lambda$ to be an adjoint symmetry for (\ref{G-Eeq-1}),
whereas the second condition (\ref{self-adjoint}) requires $\Lambda$ to be a self-adjoint function (for scalar evolution equations this means even-order).
%(i.e. $\Lambda$ must be even-order).
Here $L_E^*[u]$ denotes the adjoint operator of $L_E[u]$, namely
\begin{gather}
L_E^*[u]:=\pde{E}{u}-D_t\circ \pde{E}{u_t}-D_x\circ \pde{E}{u_x}+D_x^2\circ \pde{E}{u_{xx}}
-D_x^3\circ \pde{E}{u_{3x}}+\cdots.
\end{gather}

\strut\hfill

\noindent
The condition on the recursion operator $R[u]$ for (\ref{G-Eeq-1}) is
\begin{gather}
\label{Recursion-Condition}
[L_F[u],\,R[u]]\varphi =(D_tR[u])\varphi,
\end{gather}
where $[\ ,\ ]$ denotes the commutator (or Lie bracket). Condition (\ref{Recursion-Condition}) is evaluated on the equation (\ref{G-Eeq-1}).
Moreover, the recursion operator of (\ref{G-Eeq-1})
should generate a hierarchy of symmetries coefficients $\eta$ 
for (\ref{G-Eeq-1}), 
i.e. symmetry generators of the form
\begin{gather}
Z=\eta[u]\pde{\ }{u},
\end{gather}
by acting $R[u]$ repeatedly on $\eta$. That is
\begin{gather}
R^k[u] \eta_1[u]=\eta_{k+1}[u],\qquad k=1,2,\ldots.
\end{gather}
%Note that some of the symmetries may turn out to be nonlocal, 
%meaning that $\eta$ may contain terms of the form
%\begin{gather}
%\int g(x,u)\,dx.
%\end{gather}
For a symmetry-integrable evolution equation we require that all symmetries coefficients $\eta$ generated by $R$ are local, so a recursion operator for that equation 
would generate a hierarchy of local evolution equations 
\begin{gather}
\label{Hier-G}
u_{t_k}=R^k[u] F[u],\qquad k=1,2,\ldots.
\end{gather}
Each evolution equation in the hierarchy (\ref{Hier-G}) should
share the same set of symmetries that are generated by acting the recursion operator on the first  (or seed) member
for the hierarchy of (\ref{G-Eeq-1}). Those symmetries then span an 
Abelian Lie algebra and the recursion operator is 
hereditary for each member of the hierarchy (see \cite{Fokas-Fuchss} and \cite{olver-book} for more details).

\section{Recursion operators for a class of third-order symmetry-integrable equations}

Our starting point is the general class of third-order autonomous evolution equations of the form
\begin{gather}
\label{3rd-order-gen}
E:=u_t-F(u_x,u_{xx},u_{xxx})=0.
\end{gather}

%\noindent
%{\remark 
%Note that we have dropped the explicit dependence on $u$ in the equation (\ref{3rd-order-gen}). This of course restricts the class of equations. 
%A more general class will be considered in future (see the section {\it ``Concluding remarks''} at the end of this paper).}
%\strut\hfill
 
 \noindent
For the symmetry-integrability of (\ref{3rd-order-gen}) we need to establish a recursion operator for the equation. In this paper we consider
second-order recursion operators $R[u]$ of the form
\begin{gather}
\label{R-2nd-order-gen}
R[u]= 
G_2[u]D_x^2 +G_1[u]D_x+G_0[u]+
I_1[u]D_x^{-1}\circ \Lambda_1[u]
+
I_2[u]D_x^{-1}\circ \Lambda_2[u].
\end{gather}
%Here $G_j[u]$ depends on $u$ and its $x$-derivatives. 
The explicit conditions on $G_j$, $I_j$, $\Lambda_j$ and $F$ for (\ref{3rd-order-gen}) are given in Appendix A.

\strut\hfill

In order to find equations of the form (\ref{3rd-order-gen}) that may admit a recursion operator of the form 
 (\ref{R-2nd-order-gen}), we first establish the most general form of $F$ in terms of it highest derivative $u_{xxx}$. This is achieved by 
 solving the first three equations in the split commutator condition (\ref{Recursion-Condition}), namely those conditions on $G_j$, and $F$ that do not involve the conditions on the integrating factors $\Lambda_j$ or the symmetries $I_j$. These are the conditions (\ref{R-Cond-1}), (\ref{R-Cond-2}) and (\ref{R-Cond-3}) given in Appendix A.

{\proposition
\label{Proposition 1}
In terms of the variable $u_{xxx}$,
the most general form of $F(u_x,u_{xx},u_{xxx})$ for which (\ref{3rd-order-gen}) admits a recursion operator of the form (\ref{R-2nd-order-gen}) is given by the following four cases:
\begin{subequations}
\begin{gather}
\label{F1-gen}
F=\frac{Q_3(u_x,u_{xx})
\left[
\vphantom{\frac{DA}{DB}}
u_{xxx}+Q_2(u_x,u_{xx})\right]}
{Q_1(u_x,u_{xx})\left[
\vphantom{\frac{DA}{DB}}
Q_1(u_x,u_{xx})+(u_{xxx}+Q_2(u_x,u_{xx}))^2\right]^{1/2}}
+Q_4(u_x,u_{xx})
\\[0.9mm]
\label{F2-gen}
F=Q_1(u_{x},u_{xx})\, u_{xxx}+Q_2(u_{x},u_{xx})\\[0.9mm]
\label{F3-gen}
F=\frac{Q_1(u_x,u_{xx})}{\left[
\vphantom{\frac{DA}{DB}}
u_{xxx}+Q_2(u_x,u_{xx})\right]^2}+Q_3(u_x,u_{xx})\\[0.4cm]
\label{F4-gen}
F=\frac{Q_1(u_x,u_{xx})(u_{xxx}+Q_2(u_x,u_{xx}))}
{Q_2^2(u_x,u_{xx}) 
\left[
\vphantom{\frac{DA}{DB}}
u_{xxx}^2+2Q_2(u_x,u_{xx})u_{xxx}\right]^{1/2}  }+Q_3(u_x,u_{xx}).
\end{gather}
\end{subequations}
The functions $Q_1,\ Q_2,\ Q_3$ and $Q_4$ are arbitrary in their indicated arguments.
}

\strut\hfill

\noindent
{\bf Proof:} Solving (\ref{R-Cond-1}), (\ref{R-Cond-2}) and (\ref{R-Cond-3}) we obtain the following condition on $F(u_x,u_{xx},$ $u_{xxx})$:
\begin{gather}
\label{Cond-F}
F'\left[9(F')^2F^{(4)}
-45F'F''F'''+40(F'')^3\right]=0,
\end{gather}
where the primes denote partial derivatives with repect to $u_{xxx}$ and $F^{(4)}$ the fourth partial derivative with repect to $u_{xxx}$.
The general solution of (\ref{Cond-F}) is given by (\ref{F1-gen}), whereby (\ref{F2-gen}), (\ref{F3-gen}) and (\ref{F4-gen}) are singular solutions.
\strut\hfill$\Box$

\strut\hfill

\noindent
{\remark
We remark that the conditions given in Proposition \ref{Proposition 1} are consistent with the conditions (2.3), (2.4) and (2.5) reported in 
\cite{H-H-2005}.}

\strut\hfill

\noindent
The functions $Q_1,\ Q_2,\ Q_3$ and $Q_4$ should now be determined to gain recursion operators of the form (\ref{R-nth-order}) for the equation
(\ref{3rd-order-gen}) for each case $F$ listed in Proposition 1. This identifies the exact form of $F$ for the
symmetry-integrability of (\ref{3rd-order-gen}), which is achieved by solving the remaining conditions (\ref{R-Cond-4}), (\ref{R-Cond-5}), (\ref{R-Cond-6}) and (\ref{R-Cond-7}) given in Appendix A.

In the current paper we restrict ourselves to the case where $F(u_x,u_{xx},u_{xxx})$ is a rational functions in $u_{xxx}$, namely case (\ref{F3-gen}). This leads to the following 

\strut\hfill

\noindent
{\proposition
\label{Proposition 2}
The following equations, in the class $u_t=F(u_x,u_{xx},u_{xxx})$ with $F$ a rational function in $u_{xxx}$, are
symmetry-integrable:
\begin{itemize}
\item
%%%%%%%%%%%%%%%%%%%%%%%%
{\bf Case I}
\begin{subequations}
\begin{gather}
\label{Case-I-eq}
u_t=\frac{u_{xx}^6}{(\alpha u_x+\beta)^3u_{xxx}^2}+Q(u_x), 
\end{gather}
where $\{\alpha,\ \beta\}$ are arbitrary constants, not simultaneously zero, and $Q(u_x)$ needs to satisfy
\begin{gather}
\label{Case1-Q-cond}
(\alpha u_x+\beta)\frac{d^5Q}{du_x^5}+5\alpha \frac{d^4Q}{du_x^4}=0,
\end{gather}
which admits for $\alpha\neq 0$ the general solution
\begin{gather}
Q(u_x)=c_5\left(u_x+\frac{\beta}{\alpha}\right)^3+c_4\left(u_x+\frac{\beta}{\alpha}\right)^2
+c_3\left(u_x+\frac{\beta}{\alpha}\right)\nn\\[0.3cm]
\qquad 
+c_2\left(u_x+\frac{\beta}{\alpha}\right)^{-1}
+c_1.
\end{gather}
For $\alpha=0$, the general solution of (\ref{Case1-Q-cond}) is
\begin{gather}
Q(u_x)=c_5u_x^4+c_4u_x^3+c_3u_x^2+c_2u_x+c_1.
\end{gather}
\end{subequations}
Here $c_j$ are constants of integration.

%%%%%%%%%%%%%%%%%%%%%%%%
\item
{\bf Case II}
\begin{gather}
\label{Case-II-eq}
u_t=\frac{u_{xx}^3\left(\lambda_1+\lambda_2 u_{xx}\right)^3}{u_{xxx}^2},
\end{gather}
where $\{\lambda_1,\ \lambda_2\}$ are arbitrary constants but not simultaneously zero. 
%%%%%%%%%%%%%%%%%%%%%%%
\item
{\bf Case III}
\begin{gather}
\label{Case-III-eq}
u_t=\frac{(\alpha u_x+\beta)^{11}}{\left[
\vphantom{\frac{DA}{DB}}
(\alpha u_x+\beta)u_{xxx}-3\alpha u_{xx}^2\right]^2},
\end{gather}
where $\{\alpha,\ \beta\}$ are arbitrary constants but not simultaneously zero.
%%%%%%%%%%%%
%%%File Name for Case II
%2022-October-27-reducrsion-operator-order-2-3rd-fiully-nonlinear-F(ux,uxx,uxxx)-Case3.B-Q3(ux,uxx)=Q33(ux)-Q5(ux)-linear-k6-not-0-Case1.A.a
%%%%%%%%%%%%%%%%%%%%%%%
\item
{\bf Case IV}
\begin{gather}
\label{Case-IV-eq}
%u_t=\frac{4u_x^5}{(2u_xu_{xxx}-3u_{xx}^2)^2}\equiv \frac{u_x}{S^2},
u_t=\frac{4u_x^5}{(2b\,u_x^2-2u_xu_{xxx}+3u_{xx}^2)^2}\equiv \frac{u_x}{(b-S)^2},
\end{gather}
where $b$ is an arbitrary constant and $S$ is the Schwarzian derivative (\ref{Schwarzian}).

\end{itemize}
}

\strut\hfill

\noindent
The recursion operators for each equation listed in Proposition \ref{Proposition 2} have been computed and are given in Appendix B. 
Note that equation (\ref{Case-IV-eq}) is identical to the Möbius-invariant equation (\ref{S-Eq-1}). This recursion operator for equation 
(\ref{S-Eq-2}) is obtained by setting $b=0$ in the recursion operator (\ref{Case-IV-R}) of (\ref{Case-IV-eq}). 

\strut\hfill

For each equation listed in Proposition \ref{Proposition 2} one can easily remove the nonlinearity in the third derivative by a simple substitution $u_x=W(x,t)$
which, in a sense, ``unpotentialises'' the equations of Proposition \ref{Proposition 2}. 
For completeness, we list the so obtained equations here:

\begin{itemize}
\item
Case I:
With $u_x=W(x,t)$, (\ref{Case-I-eq}) takes the form
\begin{subequations}
\begin{gather}
W_t=-\frac{2W_x^6W_{xxx}}{(\alpha W+\beta)^3W_{xx}^3}
-\frac{3\alpha W_x^7}{(\alpha W+\beta)^4W_{xx}^2}
+\frac{6W_x^5}{(\alpha W+\beta)^3W_{xx}}\nn\\[0.3cm]
\label{Case-I-UP}
\qquad
+Q'(W)W_x,
\end{gather}
where 
\begin{gather}
(\alpha W+\beta)Q^{(5)}+5\alpha Q^{(4)}=0,\quad Q=Q(W).
\end{gather}
\end{subequations}

\item
Case II:
With $u_x=W_1(x,t)$, we obtain for (\ref{Case-II-eq}) the following equation:
\begin{gather}
\label{Case-II-W1}
W_{1,t}=
-\frac{2W_{1,x}^3(\lambda_1+\lambda_2 W_{1,x})^3W_{1,xxx}}{W_{1,xx}^3}
+\frac{3\lambda_2 W_{1,x}^3(\lambda_1+\lambda_2 W_{1,x})^2}{W_{1,xx}}\nn\\[0.3cm]
\qquad
+\frac{3W_{1,x}^2(\lambda_1+\lambda_2 W_{1,x})^3}{W_{1,xx}}.
\end{gather}
With $W_{1,x}=W_2(x,t)$, we obtain for (\ref{Case-II-W1}) the following equation:
\begin{gather}
%\label{Case-II-W2}
W_{2,t}=
-\frac{2W_2^2(\lambda_1+\lambda_2 W_2)^3W_{2,xxx}}{W_{2,x}^3}
+\frac{6W_2^3(\lambda_1+\lambda_2 W_2)^3W_{2,xx}^2}{W_{2,x}^4}\nn\\[0.3cm]
\qquad
-\frac{9W_2^2(\lambda_1+\lambda_2 W_2)^3(W_2+1)W_{2,xx}}{W_{2,x}^2}
%-\frac{9W_2^2(\lambda_1+\lambda_2 W_2)^3W_{2,xx}}{W_{2,x}^2}
+\frac{9\lambda_2W_2^2(\lambda_1+\lambda_2 W_2)^2(W_{2,x}+1)}{W_{2,x}}\nn\\[0.3cm]   
\label{Case-II-W2}
\qquad
-6\lambda_2^2W_2^3(\lambda_1+\lambda_2 W_2)
+6W_2(\lambda_1+\lambda_2 W_2)^3.
\end{gather}

\item
Case III:
With $u_x=W(x,t)$, we obtain for (\ref{Case-III-eq}) the following equation:
\begin{gather}
W_t=-\frac{(\alpha W+\beta)^{10}}{
\left[
\vphantom{\frac{DA}{DB}}
(\alpha v+\beta)W_{xx}-3\alpha W_x^2\right]^3}
\left[
\vphantom{\frac{DA}{DB}}
2\alpha WW_{xxx}\left(\alpha W+2\beta\right)
+2\beta^2W_{xxx}\right. \nn\\[0.3cm]
\label{Case-III-UP}
\qquad\qquad
\left.
\vphantom{\frac{DA}{B}}
-21\alpha W_xW_{xx}\left(\alpha W+\beta\right)
+33\alpha^2W_x^3\right].
\end{gather}

\item
Case IV:
With $u_x=W(x,t)$, we obtain for (\ref{Case-IV-eq}) the following equation:
\begin{gather}
W_t=\frac{4W^4}{
\left(
\vphantom{\frac{DA}{DB}}
2bW^2-2WW_{xx}+3W_x^2\right)^3}
\left(
\vphantom{\frac{DA}{DB}}
4W^2W_{xxx}-18WW_xW_{xx}
\right.\nn\\[0.3cm]
\label{Case-IV-UP}
\qquad\qquad
\left.
\vphantom{\frac{DA}{DB}}
+15W_x^3+2bW^2W_{x}\right).
\end{gather}

\end{itemize}

\section{Concluding remarks}
Our aim has been to construct fully-nonlinear third-order evolution equations in the class $u_t=F(u_x,u_{xx},u_{xxx})$, namely to identify those equations in this class 
that admit a second-order recursion operator with a sixth-order integrating factor, which are then symmetry-integrable equations. 
Note that that exists no fully-nonlinear evolution equation in this class that admits a recursion operator of order two where both integrating factors, $\Lambda_1$ and $\Lambda_2$, are of order less than six.

We report here four equations, listed in Proposition 2, namely (\ref{Case-I-eq}),  (\ref{Case-II-eq}), (\ref{Case-III-eq}) and (\ref{Case-IV-eq}). Due to the mentioned restrictions on the form of the recursion operator, this is certainly not a complete classification of all fully-nonlinear third-order evolution equations of this form
that admit a recursion operator. Nevertheless, we do consider the equations that we have obtained here to be of some
interest and worthy of further study. It would, for example, be interesting to find all the potentialisations of the four fully-nonlinear equations 
(\ref{Case-I-eq}) to (\ref{Case-IV-eq}), as well as the equations (\ref{Case-I-UP}) to (\ref{Case-IV-UP}).
This can be investigated by using the adjoint symmetries structure of the equations. Some preliminary calculations have revealed a rich adjoint symmetry structure for these equations, so one can expect to obtain interesting results. Furthermore, one could apply the multi-potentialisation method which may lead to nonlocal symmetries for the equations 
(see \cite{Euler-Euler-book-article} for details regarding multi-potentialisations). 
One could also extend this study further, namely to include evolution equations of third order that explicitly depend on $u$ and allow algebraic functions in $u_{xxx}$.

%These will be the topics of some future papers.

\subsection*{Acknowledgements}
We thank the anonymous referee for useful remarks regarding the classification of integrable evolution equations, for point out some misprints, and for clarifying the results in \cite{H-H-2005}.

\renewcommand{\theequation}{\thesection.\arabic{equation}}
\setcounter{equation}{0}

\appendix 
\section{Appendix: The general conditions for $R[u]$ of  (\ref{3rd-order-gen}) }
For the equation $u_t=F(u_x,u_{xx},u_{xxx})$ we provide here the explicit general conditions on the functions $F$, $G_j$, $I_j$ and $\Lambda_j$ for the existence of a recursion operator $R[u]$ of the form 
\begin{gather}
%\label{R-2nd-order-gen}
R[u]= 
G_2[u]D_x^2 +G_1[u]D_x+G_0[u]+
I_1[u]D_x^{-1}\circ \Lambda_1[u]
+
I_2[u]D_x^{-1}\circ \Lambda_2[u].
\end{gather}
This is obtained from the commutator condition
(\ref{Recursion-Condition})
by equating to zero all the derivatives of the free function $\varphi$.
For convenience we introduce the following notation: 
\begin{gather*}
A_1:=\pde{F}{u_x},\qquad
A_2:=\pde{F}{u_{xx}},\qquad
A_3:=\pde{F}{u_{xxx}}.
\end{gather*}
The conditions are as follows:
\begin{subequations}
\begin{gather}
%%%%%%%%%%%%%%%%%\label{R-Cond-1}
\label{R-Cond-1}
\frac{\p^4\varphi}{\p x^4}:
-2G_2D_x A_3+3A_3D_x G_2=0\\[0.3cm]
%%%%%%%%%%%%%%%%%\label{R-Cond-2}
%\label{R-Cond-2}
\frac{\p^3\varphi}{\p x^3}:
2A_2D_xG_2
-2G_2D_xA_2
-G_2D_x^2A_3
-G_1D_xA_3
+3A_3D_x^2G_2\nn\\[0.3cm]
\label{R-Cond-2}
\qquad\ 
+3A_3D_xG_1=0\\[0.3cm]
%%%%%%%%%%%%%%%\label{R-Cond-3}
\frac{\p^2\varphi}{\p x^2}:
3A_3D_x^2G_1
+A_3D_x^3G_2
+2A_2D_xG_1
+A_2D_x^2G_2
+A_1D_xG_2
+3A_3D_xG_0\nn\\[0.3cm]
\label{R-Cond-3}
\qquad\ 
\left.
\vphantom{\frac{DA}{DB}}
-2G_2D_xA_1
-G_2D_x^2A_2
-G_1D_xA_2
-D_tG_2\right|_{E=0}
=0\\[0.3cm]
%%%%%%%%%%%%%%\label{R-Cond-4}
\frac{\p\varphi}{\p x}:
A_3D_x^3G_1
+3A_3D_x^2G_0
+A_2D_x^2G_1
+2A_2D_xG_0
+A_1D_xG_1\nn\\[0.3cm]
\qquad\ 
+\sum_{j=1}^2\left(
\vphantom{\frac{DA}{DB}}
3A_3\Lambda_jD_xI_j
+3A_3I_jD_x\Lambda_j
+I_j\Lambda_jD_xA_3\right)\nn\\[0.3cm]
\label{R-Cond-4}
\qquad\ 
\left.
\vphantom{\frac{DA}{DB}}
-G_2D_x^2A_1
-G_1D_xA_1
-D_tG_1\right|_{E=0}
=0\\[0.3cm]
%%%%%%%%%%%%%\label{R-Cond-5}
\varphi:
A_3D_x^3G_0
+A_2D_x^2G_0
+A_1D_xG_0
+\sum_{j=1}^2\left(
\vphantom{\frac{DA}{DB}}
-2I_j(D_x\Lambda_j)(D_xA_3)
-I_j\Lambda_jD_x^2A_3\right.
\nn\\[0.3cm]
\qquad\ 
\left.
+I_j\Lambda_jD_xA_2
-I_jD_x^4\Lambda_j
+3A_3(D_xI_j)(D_x\Lambda_j)
+3A_3\Lambda_jD_x^2I_j\right.
\nn\\[0.3cm]
\label{R-Cond-5}
\qquad\ 
\left.\left.
\vphantom{\frac{DA}{DB}}
+2A_2\Lambda_jD_xI_j
+2A_2I_jD_x\Lambda_j
\right)
-D_tG_0\right|_{E=0}=0,
\end{gather}
%\end{subequations}
as well as the symmetry condition 
\begin{gather}
\label{R-Cond-6}
\left.
\vphantom{\frac{DA}{DB}}
L_E[u]\,I_j\right|_{E=0}=0,\qquad j=1,2
\end{gather}
and the adjoint symmetry condition
\begin{gather}
\label{R-Cond-7}
\left.\vphantom{\frac{DA}{DB}}
L^*_E[u]\, \Lambda_j\right|_{E=0}=0,\qquad j=1,2.
\end{gather}
\end{subequations}

\section{Appendix: The recursion operators for the symmetry-integrable equations of Proposition \ref{Proposition 2}}

\noindent
{\bf Recursion operator for Case I:}
Equation (\ref{Case-I-eq}) of Proposition \ref{Proposition 2} viz.
\begin{gather*}
%\label{Case-I-eq}
u_t=\frac{u_{xx}^6}{(\alpha u_x+\beta)^3u_{xxx}^2}+Q(u_x),
\end{gather*}
admits the
recursion operator 
\begin{gather}
R[u]=G_2[u]D_x^2+G_1[u]D_x+G_0[u]+(\alpha u_x+\beta)D_x^{-1}\circ \Lambda_1[u],
\end{gather}
where
\begin{subequations}
\begin{gather}
G_2[u]=\frac{u_{xx}}{(\alpha u_x+\beta)^2\,u_{xxx}^2}\\[0.3cm]
G_1[u]=\frac{u_{xx}^4u_{4x}}{(\alpha u_x+\beta)^2\,u_{xxx}^3}
-\frac{4u_{xx}^3}{(\alpha u_x+\beta)^2\,u_{xxx}}
+\frac{\alpha u_{xx}^5}{(\alpha u_x+\beta)^3\,u_{xxx}^2}\\[0.3cm]
G_0[u]=-\frac{u_{xx}^4u_{5x}}{(\alpha u_x+\beta)^2\,u_{xxx}^3}
+\frac{3u_{xx}^4u_{4x}^2}{(\alpha u_x+\beta)^2\,u_{xxx}^4}\nn\\[0.3cm]
\qquad
+\left(-\frac{8u_{xx}^3}{(\alpha u_x+\beta)^2u_{xxx}^2}
+\frac{6\alpha u_{xx}^5}{(\alpha u_x+\beta)^3\,u_{xxx}^3}\right)u_{4x}
+\frac{6\alpha^2u_{xx}^6}{(\alpha u_x+\beta)^4\,u_{xxx}^2}\nn\\[0.3cm]
\qquad
-\frac{18\alpha u_{xx}^4}{(\alpha u_x+\beta)^3\,u_{xxx}}
+\frac{12\alpha u_{xx}^2}{(\alpha u_x+\beta)^2}
+\frac{1}{3}(\alpha u_x+\beta)\frac{d^2Q}{du_x^2}
-\frac{\alpha}{3}\frac{dQ}{du_x}\\[0.3cm]
\Lambda_1=\frac{u_{xx}^4u_{6x}}{(\alpha u_x+\beta)^3\,u_{xxx}^3}
+\left(
\frac{12u_{xx}^3}{(\alpha u_x+\beta)^3\,u_{xxx}^2}
-\frac{9\alpha u_{xx}^5}{(\alpha u_x+\beta)^4\,u_{xxx}^3}\right)u_{5x}
\nn\\[0.3cm]
\qquad
+\left(
\frac{24\alpha u_{xx}^2}{(\alpha u_x+\beta)^3\,u_{xxx}}
-\frac{72\alpha u_{xx}^4}{(\alpha u_x+\beta)^4\,u_{xxx}^2}
+\frac{36\alpha^2 u_{xx}^6}{(\alpha u_x+\beta)^5\,u_{xxx}^3}
\right)u_{4x}
\nn\\[0.3cm]
\qquad
-\frac{9u_{xx}^4u_{4x}u_{5x}}{(\alpha u_x+\beta)^3\,u_{xxx}^4}
+\left(
\frac{27\alpha u_{xx}^5}{(\alpha u_x+\beta)^4\,u_{xxx}^4}
-\frac{28u_{xx}^3}{(\alpha u_x+\beta)^3\,u_{xxx}^3}
\right)u_{4x}^2
\nn\\[0.3cm]
\qquad
+\frac{12u_{xx}^4u_{4x}^3}{(\alpha u_x+\beta)^3\,u_{xxx}^5}
-\frac{24u_{xx}u_{xxx}}{(\alpha u_x+\beta)^3}
+\frac{30\alpha^3u_{xx}^7}{(\alpha u_x+\beta)^6\,u_{xxx}^2}
+\frac{108\alpha u_{xx}^3}{(\alpha u_x+\beta)^4}
\nn\\[0.3cm]
\qquad
-\frac{108\alpha^2u_{xx}^5}{(\alpha u_x+\beta)^5\,u_{xxx}}
-\frac{1}{2}\frac{d^3Q}{du_x^3}\,u_{xx}.
\end{gather}
\end{subequations}
Here $Q(u_x)$ needs to satisfy the 5th-order ordinary differential equation (\ref{Case1-Q-cond}), viz.
\begin{gather*}
%\label{Case1-Q-cond}
(\alpha u_x+\beta)\frac{d^5Q}{du_x^5}+5\alpha \frac{d^4Q}{du_x^4}=0.
\end{gather*}

\strut\hfill

\noindent
{\bf Recursion operator for Case II:}
Equation (\ref{Case-II-eq}) of Proposition \ref{Proposition 2} viz.
\begin{gather*}
%\label{Case-II-eq}
u_t=\frac{u_{xx}^3\left(\lambda_1+\lambda_2 u_{xx}\right)^3}{u_{xxx}^2}
\end{gather*}
%
%\begin{gather*}
%\label{Case-III-eq}
%u_t=\frac{4u_x^5}{(2u_xu_{xxx}-3u_{xx}^2)^2}.
%\end{gather*}
admits the
recursion operator 
\begin{gather}
R[u]=G_2[u]D_x^2+G_1[u]D_x+G_0[u]+D_x^{-1}\circ \Lambda_1[u],
\end{gather}
where
\begin{subequations}
\begin{gather}
G_2[u]=\frac{u_{xx}^2(\lambda_1+\lambda_2u_{xx})^2}{u_{xxx}^2}\\[0.3cm]
G_1[u]=\frac{u_{xx}^2(\lambda_1+\lambda_2u_{xx})^2 u_{4x}}{u_{xxx}^3}
-\frac{4\lambda_2u_{xx}^2(\lambda_1+\lambda_2u_{xx})}{u_{xxx}}
\\[0.3cm]
G_0[u]=\frac{u_{xx}^2(\lambda_1+\lambda_2u_{xx})^2 u_{5x}}{u_{xxx}^3}
+\frac{3u_{xx}^2(\lambda_1+\lambda_2u_{xx})^2 u_{4x}^2}{u_{xxx}^4}
\nn\\[0.3cm]
\qquad
-\frac{2u_{xx}u_{xxx}^2(\lambda_1+\lambda_2u_{xx})(\lambda_1+4\lambda_2u_{xx})   u_{4x}}{u_{xxx}^4}
+12\lambda_2^2u_{xx}^2
+6\lambda_1\lambda_2u_{xx}\\[0.3cm]
\Lambda_1[u]=\frac{u_{xx}^2(\lambda_1+\lambda_2u_{xx})^2u_{6x}}{u_{xxx}^3}
+\frac{4u_{xx}(\lambda_1+\lambda_2u_{xx})(\lambda_1+3\lambda_2u_{xx})u_{5x}}{u_{xxx}^2}
\nn\\[0.3cm]
\qquad
-\frac{9u_{xx}^2(\lambda_1+\lambda_2u_{xx})^2u_{4x}u_{5x} }{u_{xxx}^4}
+\frac{12u_{xx}^2(\lambda_1+\lambda_2u_{xx})^2u_{4x}^3}{u_{xxx}^5}
\nn\\[0.3cm]
\qquad
-\frac{2u_{xx}u_{xxx}^3(\lambda_1+\lambda_2u_{xx})(5\lambda_1+14\lambda_2u_{xx})u_{4x}^2}{u_{xxx}^5}
\nn\\[0.3cm]
\qquad
+\frac{2(12\lambda_2^2u_{xx}^2+10\lambda_1\lambda_2u_{xx}+\lambda_1^2)u_{4x}}{u_{xxx}}
-6\lambda_2(\lambda_1+4\lambda_2u_{xx})u_{xxx}.
\end{gather}
\end{subequations}

\strut\hfill

\noindent
{\bf Recursion operator for Case III:}
Equation (\ref{Case-III-eq}) of Proposition \ref{Proposition 2} viz.
\begin{gather*}
%\label{Case-III-eq}
u_t=\frac{(\alpha u_x+\beta)^{11}}{\left[
\vphantom{\frac{DA}{DB}}
(\alpha u_x+\beta)u_{xxx}-3\alpha u_{xx}^2\right]^2}
\end{gather*}
admits the
recursion operator 
\begin{gather}
R[u]=G_2[u]D_x^2+G_1[u]D_x+G_0[u]+(\alpha u_x+\beta)D_x^{-1}\circ \Lambda_1[u]
\end{gather}
where
\begin{subequations}
\begin{gather}
G_2[u]=
\frac{(\alpha u_x+\beta)^8}{\left[
\vphantom{\frac{DA}{DB}}
(\alpha u_x+\beta)u_{xxx}-3\alpha u_{xx}^2\right]^2}\\[0.3cm]
G_1[u]=
\frac{(\alpha u_x+\beta)^7\,u_{4x}}{\left[
\vphantom{\frac{DA}{DB}}
(\alpha u_x+\beta)u_{xxx}-3\alpha u_{xx}^2\right]^3}
\left[
\vphantom{\frac{DA}{DB}}
(\alpha u_x+\beta)^2u_{4x}
-13\alpha(\alpha u_x+\beta)u_{xx}u_{xxx}\right.\nn\\[0.3cm]
\qquad\qquad
\left.
\vphantom{\frac{DA}{DB}}
+24\alpha^2u_{xx}^3\right]\\[0.3cm]
G_0[u]=
-\frac{(\alpha u_x+\beta)^9u_{5x}}{\left[
\vphantom{\frac{DA}{DB}}
(\alpha u_x+\beta)u_{xxx}-3\alpha u_{xx}^2\right]^3}\nn\\[0.3cm]
\qquad
+\frac{3(\alpha u_x+\beta)^6}{\left[
\vphantom{\frac{DA}{DB}}
(\alpha u_x+\beta)u_{xxx}-3\alpha u_{xx}^2\right]^4} 
\left[
\vphantom{\frac{DA}{DB}}
(\alpha u_x+\beta)^4u_{4x}^2
-\frac{46}{3}\alpha(\alpha u_x+\beta)^3u_{xx}u_{xxx}u_{4x}
\right.\nn\\[0.3cm]
\qquad\qquad\ 
+3\alpha(\alpha u_x+\beta)^3u_{xxx}^3
+\frac{184}{3}\alpha^2(\alpha u_x+\beta)^2
u_{xx}^2u_{xxx}^2\nn\\[0.3cm]
\qquad\qquad
\left.
\vphantom{\frac{DA}{DB}}
-184\alpha^3(\alpha u_x+\beta)u_{xx}^4u_{xxx}
+144\alpha^4u_{xx}^6\right]\\[0.3cm]
\Lambda_1[u]=
\frac{(\alpha u_x+\beta)^8\,u_{6x}}
{\left[
\vphantom{\frac{DA}{DB}}
(\alpha u_x+\beta)u_{xxx}-3\alpha u_{xx}^2\right]^3}
-\frac{9(\alpha u_x+\beta)^9\,u_{4x}u_{5x}}
{\left[
\vphantom{\frac{DA}{DB}}
(\alpha u_x+\beta)u_{xxx}-3\alpha u_{xx}^2\right]^4}\nn\\[0.3cm]
\qquad
-\frac{72\alpha^2(\alpha u_x+\beta)^7\,u_{xx}^3u_{5x}}
{\left[
\vphantom{\frac{DA}{DB}}
(\alpha u_x+\beta)u_{xxx}-3\alpha u_{xx}^2\right]^4}
+\frac{81\alpha(\alpha u_x+\beta)^8\,u_{xx}u_{xxx}u_{5x}}
{\left[
\vphantom{\frac{DA}{DB}}
(\alpha u_x+\beta)u_{xxx}-3\alpha u_{xx}^2\right]^4}\nn\\[0.3cm]
\qquad
+\frac{12(\alpha u_x+\beta)^{10}\,u_{4x}^3}
{\left[
\vphantom{\frac{DA}{DB}}
(\alpha u_x+\beta)u_{xxx}-3\alpha u_{xx}^2\right]^5}\nn\\[0.3cm]
\qquad
-\frac{45\alpha(\alpha u_x+\beta)^8\,u_{xx}u_{4x}^2}
{\left[
\vphantom{\frac{DA}{DB}}
(\alpha u_x+\beta)u_{xxx}-3\alpha u_{xx}^2\right]^5}
\left(5\alpha u_xu_{xxx}+5\beta u_{xxx}-3\alpha u_{xx}^2\right)\nn\\[0.3cm]
\qquad
+\frac{5\alpha(\alpha u_x+\beta)^6\,u_{4x}}
{\left[
\vphantom{\frac{DA}{DB}}
(\alpha u_x+\beta)u_{xxx}-3\alpha u_{xx}^2\right]^5}
\left[
\vphantom{\frac{DA}{DB}}
11(\alpha u_x+\beta)^3u_{xxx}^3
+291\alpha (\alpha u_x+\beta)^2u_{xx}^2u_{xxx}^2\right.\nn\\[0.3cm]
\qquad\qquad
\left.
\vphantom{\frac{DA}{DB}}
-504\alpha^2(\alpha u_x+\beta)u_{xx}^4u_{xxx}
+216\alpha^3u_{xx}^6\right]\nn\\[0.3cm]
\qquad
-\frac{20\alpha^2(\alpha u_x+\beta)^4\,u_{xx}}
{\left[
\vphantom{\frac{DA}{DB}}
(\alpha u_x+\beta)u_{xxx}-3\alpha u_{xx}^2\right]^7}
\left[
\vphantom{\frac{DA}{DB}}
-\frac{67\alpha^2}{3}(\alpha u_x+\beta)^4u_{xx}^4u_{xxx}^4\right.\nn\\[0.3cm]
\qquad\qquad
+148\alpha^3(\alpha u_x+\beta)^3 u_{xx}^6 u_{xxx}^3
+288\alpha^5(\alpha u_x+\beta)u_{xx}^{10}u_{xxx}\nn\\[0.3cm]
\qquad\qquad
-306\alpha^4(\alpha u_x+\beta)^2u_{xx}^8u_{xxx}^2
+\frac{31}{27}(\alpha u_x+\beta)^6u_{xxx}^6\nn\\[0.3cm]
\qquad\quad\ \,
\left.
\vphantom{\frac{DA}{DB}}
-\frac{38\alpha}{9}(\alpha u_x+\beta)^5u_{xx}^2u_{xxx}^5
-108\alpha^6u_{xx}^{12}\right].
\end{gather}
\end{subequations}

\strut\hfill

\noindent
{\bf Recursion operator for Case VI:}
Equation (\ref{Case-IV-eq}) of Proposition \ref{Proposition 2} viz.
\begin{gather*}
%\label{Case-IV-eq}
%u_t=\frac{4u_x^5}{(2u_xu_{xxx}-3u_{xx}^2)^2}\equiv \frac{u_x}{S^2},
u_t=\frac{4u_x^5}{(2b\,u_x^2-2u_xu_{xxx}+3u_{xx}^2)^2}
\equiv \frac{u_x}{(b-S)^2}
,
\end{gather*}
admits the recursion operator 
\begin{gather}
\label{Case-IV-R}
R[u]=G_2[u]D_x^2+G_1[u]D_x+G_0[u]+ u_xD_x^{-1}\circ \Lambda_1[u]+u_tD_x^{-1}\circ \Lambda_2[u]
\end{gather}
where
\begin{subequations}
\begin{gather}
G_2[u]=\frac{1}{4(b-S)^2}
\\[0.3cm]
G_1[u]=-\frac{u_{xx}}{2u_x(b-S)^2}
-\frac{S_x}{4(b-S)^3}
\\[0.3cm]
G_0[u]=\frac{u_{xx}^2}{8u_x^2(b-S)^2}
+\frac{u_{xx}S_x}{4u_x(b-S)^3}
+\frac{S_{xx}}{4(b-S)^3}
-\frac{2bS^2-b^2S-3S_x^2-S^3}{4(b-S)^4}
\\[0.3cm]
\Lambda_1[u]=-\frac{S_{xxx}}{4u_x(b-S)^3}
-\frac{9S_xS_{xx}}{4u_x(b-S)^4}
-\frac{S_x(b+3S)}{8u_x(b-S)^3}
-\frac{3S_x^3}{u_x(b-S)^5}
\\[0.3cm]
\Lambda_2[u]=-\frac{S_x}{8u_x}.
\end{gather}
\end{subequations}
Here $S$ is the Schwarzian derivative (\ref{Schwarzian}).

\label{lastpage}
\end{document}